# Control of Extraordinary Optical Transmission in Resonant Terahertz Gratings via Lateral Depletion in an AlGaN/GaN Heterostructure

Geofrey Nyabere[1], Hunter Ellis[1], Miguel Gomez[1], Wei Jia[1], Yizheng Liu[2], Karli Ann Higley[1], Sriram Krishnamoorthy[2], Steve Blair[1], Kai Fu[1], and Berardi Sensale-Rodriguez[1,*]

[1] Department of Electrical and Computer Engineering, The University of Utah, Salt Lake City, UT 84112, USA

[2] Materials Department, University of California Santa Barbara, Santa Barbara, CA 93106, USA

[*] Corresponding author, e-mail: berardi.sensale@utah.edu

## Abstract

Periodic metallic gratings on substrates can support a range of electromagnetic modes, such as leaky waveguide, guided-resonant, and Fabry–Pérot (FP) cavity modes, which can strongly modulate optical transmission under resonant excitation. Here, we investigate how this coupling can be dynamically manipulated through charge-density control in a laterally patterned AlGaN/GaN heterostructure. The structure comprises metallic stripes separated by regions containing a two-dimensional electron gas (2DEG), forming a periodically modulated interface whose electromagnetic response is governed by the charge density between the stripes. In the unbiased state, the conductive 2DEG screens the incident terahertz field and suppresses excitation of guided modes. When the 2DEG is depleted, the change in boundary conditions allows efficient coupling into substrate resonances, producing a strong modulation at particular frequencies where extraordinary optical transmission (EOT) through the structure takes place. The results highlight the sensitive dependence of guided-mode-resonance (GMR) mediated EOT on inter-stripe charge distribution and demonstrate a direct interplay between carrier dynamics and resonant electromagnetic phenomena in the terahertz regime.

## I. INTRODUCTION

Active control of the electromagnetic wave propagation in the terahertz (THz) frequency range is of critical importance for applications in imaging, spectroscopy, and next-generation wireless communications. In this context, the earliest electrically driven THz modulators were reported by Kleine-Ostmann *et al* in the mid-2000s, who employed carrier depletion in a two-dimensional electron gas (2DEG) heterostructure to achieve room-temperature modulation of THz transmission[1]. Subsequent work by the same group investigated the spatial distribution of the depletion region and its effect on modulation efficiency, establishing the basic principles of semiconductor-based THz modulators[2]. Although these early devices demonstrated proof-of-concept electrical tunability, their performance was limited by the modest modulation depth and speed attainable through direct control of carrier density in large-area structures.

A significant advance was achieved with the advent of metamaterial-based modulators. Herein, Chen *et al* demonstrated the first active THz metamaterial device by integrating semiconductor layers with resonant metallic elements, enabling strong modulation of transmitted amplitude through electrical excitation[3]. Work by Shrekenhamer *et al.* subsequently incorporated high-electron-mobility transistors into metamaterial unit cells, realizing high-speed modulation with enhanced interaction between the electromagnetic resonance and the active medium[4]. These works established metamaterial approaches as an effective route to increase modulation depth by concentrating the electric field in subwavelength resonators.

The introduction of novel tunable materials further expanded the design space for active THz devices. Graphene, with its gate-tunable carrier density and high mobility, enabled modulators with near-unity amplitude modulation and low insertion loss[5,6]. Similarly, vanadium dioxide ($VO_2$) was used to realize thermal or optically driven modulators that exploit the insulator-to-metal phase transition to achieve large contrast ratios[7,8]. These developments demonstrated the potential of combining resonant structures with materials enabling dynamic conductivity control.

In parallel, periodic metallic gratings and all-dielectric metasurfaces have emerged as versatile platforms for manipulating THz transmission through guided-mode resonances (GMR) and extraordinary optical transmission (EOT). In this context, Ferraro *et al.* demonstrated narrowband THz filters based on the excitation of guided-mode resonances in thin cyclo-olefin polymer films patterned with metallic stripes or patches, achieving high-Q, high-transmission responses suitable for THz communications applications[9]. Building on this, Bark *et al.* explored all-dielectric GMR filters in the THz region and identified multiple TE and TM resonances with high Q-factors and polarization selectivity, establishing dielectric GMR filters as low-loss, tunable alternatives to metal-based metasurfaces[10]. Later work by Bark and Jeon employed the TE-mode GMR of dielectric films to realize sensitive dielectric-film sensors in the THz band[11]. Jia *et al.* extended these studies to analyze metallic gratings on anisotropic dielectric substrates, revealing the role of substrate anisotropy on the resonance characteristics and demonstrating their application for anisotropic refractive-index characterization[12].

The potential for dynamic control in GMR systems has been further demonstrated through tunable and reconfigurable designs. Miao *et al.* reported a dynamically tunable THz GMR filter fabricated on an elastic styrene–butadiene–styrene (SBS) film, achieving broadband spectral tunability through mechanical strain while preserving narrow linewidth and structural stability[13]. Furthermore, Yao *et al.* demonstrated the versatility of resonant waveguide gratings in optical sensing by integrating FP cavities with guided-mode structures, achieving enhanced resonance strength and compact footprint for refractive-index detection applications[14]. From a theoretical analysis, Isic *et al.* also showed that metal–semiconductor–metal metasurface modulators can provide substantial electrical control of resonant reflection in the THz regime[15].

Within the broader context of electrically and optically tunable terahertz modulators, Huang *et al*. demonstrated electrically controlled terahertz transmission in AlGaN/GaN heterostructures using a grating-gate–coupled two-dimensional electron gas, in which carrier depletion modulated plasmonic absorption[16]. Chen *et al*. reported electrically driven terahertz modulation in GaN-on-Si heterostructures employing patterned metallic metasurfaces integrated with a 2DEG to enable broadband and multimodal operation[17]. More recently, Sai *et al*. demonstrated optically controlled terahertz modulation in AlGaN/GaN grating-gate plasmonic crystals, exploiting nonlinear plasmonic responses[18].

Overall, these results establish that GMR and EOT-based architectures, whether metallic or all-dielectric, can simultaneously enable high spectral selectivity, strong field confinement, and tunable functionality, thereby bridging fundamental resonance phenomena with practical THz device applications.

Building upon these advances, the present work explores the control of extraordinary optical transmission in resonant THz gratings through lateral depletion in an AlGaN/GaN heterostructure. By modulating the inter-stripe carrier distribution in a two-dimensional electron gas, the boundary conditions governing guided-mode resonance coupling are dynamically altered, enabling efficient electronic tuning of resonant THz transmission.

## II. DEVICE STRUCTURE AND SIMULATIONS

The proposed device consists of an interdigitated metallic grating on a sapphire substrate, as depicted in **Fig. 1(a)**. The unit cell geometry is defined by the finger width ($w$), gap ($g$), period ($p = w + g$), metal thickness ($t$), and substrate thickness ($h$), as shown in the inset of **Fig. 1(c)**. We performed a two-dimensional full-wave electromagnetic simulation using COMSOL Multiphysics (with results verified in Ansys Electronic Desktop HFSS). The grating parameters were set as follows: $w$ =200 µm, separation gap $g$ =3 µm, metal thickness $t$ =300 nm, and substrate thickness $h$ =430 µm. The selection of a 3 µm gap serves a dual purpose: it optimizes the filling factor ($F = w/p$) for device performance while maintaining a feature size large enough to ensure reliable fabrication through optical lithography. The structure's periodicity reduced the simulation domain to a single unit cell with periodic boundary conditions defined in the $x$-direction and perfectly matched layers in the y direction. A TE-polarized plane wave (electric field oriented perpendicular to the fingers) was normally incident on the structure. The grating was modeled as gold with a finite conductivity of $\sigma_{gold} = 4.1 \times 10^7$ S/m. The substrate was modeled with 1.8 μm of GaN on the top side, with the remaining thickness composed of 426.5 μm sapphire. To illustrate the response of the structure, we analyzed two configurations: (*i*) a reference grating on a lossless sapphire substrate and (*ii*) an identical grating with a two-dimensional electron gas (2DEG) layer located just beneath the fingers. The 2DEG was modeled as a conductive sheet impedance, with its frequency-dependent permittivity and conductivity governed by a Drude model[19]. Following an approach previously used by our group[20], we fitted an analytical model to measured transmission data through a 2DEG in AlGaN/GaN on sapphire wafer to determine its charge transport parameters. From this fit, we obtained a mobility of 1,540 $cm^2/V \cdot s$, a sheet density of $1.43 \times 10^{13}$ $cm^{-2}$, and a sheet conductivity of 3.50 mS. Contrasting transmission results for (*i*) and (*ii*) enables us to predict the dynamic response of the device illustrated in **Fig. 1(b)**.

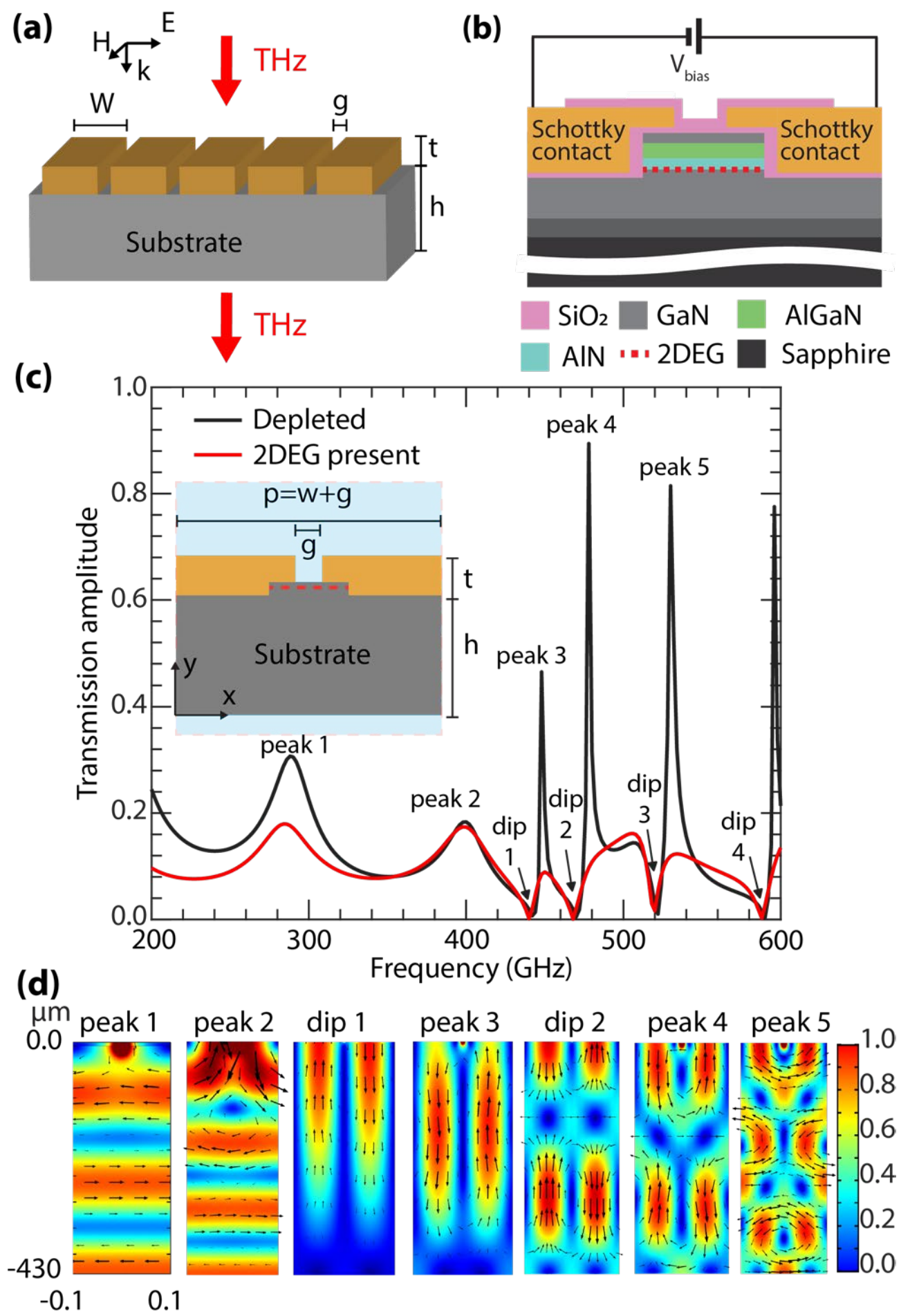

**Figure 1. (a)** Schematic of the grating structure under normally incident, TE-polarized THz radiation. **(b)** Biasing scheme for the interdigitated grating, where the applied voltage creates a depletion region to modulate the 2DEG conductivity. **(c)** The simulated transmission spectrum for a single unit cell shows characteristic FP and GMR. Spectrally narrow EOT peaks at high frequencies vanish when a 2DEG is placed in between the grating metal stripes. Inset: Unit cell used for simulation **(d)** Electric field distributions for the resonant peaks and dips labeled in (c), with arrows indicating the local field orientation.

As shown in **Fig. 1(c)**, the simulated transmission spectrum for the reference grating (*i*) shows two clear resonant regimes. At lower frequencies, the spectrum is characterized by broad Fabry-Pérot (FP) resonances within the substrate, with the N-th order mode approximated by[12],

$$f_{TEM,N} = \frac{\left(2N - \frac{w}{p}\right) c_0}{4\sqrt{\varepsilon_{xx}} h}, \qquad (1)$$

where $p$ is the grating period, $w$ is the stripe width, $\varepsilon_{xx}$ is the in-plane relative substrate permittivity (~9.4 for sapphire), and $h$ is the substrate thickness. For example, the simulated third-order FP mode at 0.288

THz is in excellent agreement with the calculated value of 0.276 THz. At higher frequencies, the response is dominated by sharp guided-mode resonances (GMRs)[9]. These arise when grating-diffracted waves couple into slab waveguide modes supported by the substrate. The general condition for GMRs is given by [12],

$$f_{TM,N_x,N_y} = c_0\sqrt{\left(\frac{N_x}{p\sqrt{\varepsilon_{yy}}}\right)^2 + \left(\frac{N_y}{2h\sqrt{\varepsilon_{xx}}}\right)^2}, \tag{2}$$

where $N$ is the integer mode order, $c_0$ is the speed of light in vacuum, $p$ is the period, $h$ is the substrate height, $\varepsilon_{xx}$ is the in-plane, and $\varepsilon_{yy}$ is the out-of-plane (~11.5 for sapphire) substrate relative permittivity. For odd order modes, ½ is added to the $N_y$ value. In our spectrum, even-order modes appear as transmission peaks (e.g., at 448, 478, and 530 GHz) due to constructive interference, while odd-order modes appear as dips (e.g., at 442, 470, 522, and 590 GHz) due to destructive interference.

Analysis of the simulated electric field distributions (**Fig. 1(d)**) confirms this behavior. The transmission peaks (3, 4) correspond to even modes with symmetric field profiles, while the dips (1, 2) correspond to odd modes with antisymmetric profiles. Similar to subwavelength slot antennas, the fields are strongly confined within the gaps and enhanced at the metal edges[9]. This strong localization is crucial for efficient coupling to the substrate modes, dictating transmission features. This observation also holds critical importance for the active operation of the device when controlling charge in the 2DEG.

When the 2DEG layer is included, the transmission spectrum changes noticeably. The grating becomes more reflective, and the sharp guided-mode resonance (GMR) peaks are strongly suppressed, as shown in **Fig. 1(c)**. The effect is more pronounced for the highest-Q resonances. At zero bias, the free carriers in the 2DEG effectively short the capacitive gaps of the grating, preventing the resonant field build-up required for efficient transmission and causing the incident THz wave to be reflected. The broader FP background is less sensitive to field confinement and changes very little. In contrast, the transmission peak at 478 GHz is strongly reduced. This shows the strong influence of the conductive channel. In an actual device, the 2DEG conductivity can be tuned with an applied voltage across the metal stripes (electrodes). The "no-2DEG" condition corresponds to a fully depleted channel, which occurs when a large enough reverse bias is applied between adjacent fingers acting as Schottky contacts (**Fig. 1(b)**). This applied voltage expands the depletion region laterally. By adjusting the applied voltage, the 2DEG can be partially or fully depleted, allowing control over the resonant THz transmission.

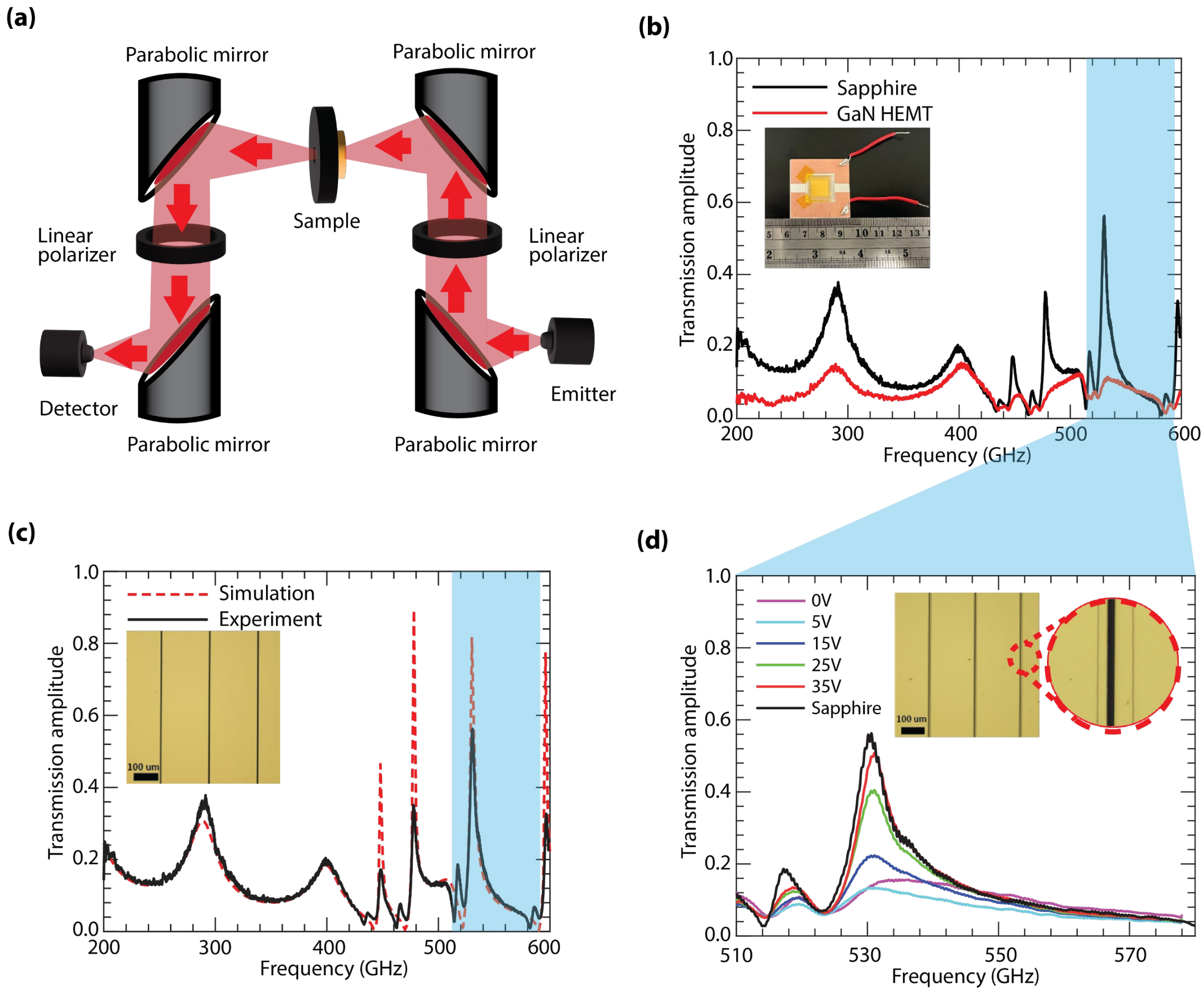

**Figure 2**. **(a)** Diagram of the frequency-domain THz spectrometer employed for device characterization. **(b)** Transmission spectra measured for the reference grating (black) and the AlGaN/GaN active device at zero bias (red). The presence of the 2DEG makes the EOT resonant features vanish from the measured spectra. Inset: microscope image of the active device. **(c)** Measured (solid) versus simulated (dashed) transmission for the passive reference grating on sapphire, showing close agreement in the position of spectral features. Inset: optical image of the fabricated reference device. **(d)** Transmission of the active device under reverse bias ranging from 0 V to 35 V. As the 2DEG is depleted, the EOT peak becomes more pronounced. Inset: zoomed-in view of the interdigitated fingers and mesa.

## III. EXPERIMENTAL RESULTS AND DISCUSSION

We fabricated two devices: an active grating on an AlGaN/GaN heterostructure and a passive reference grating on a bare sapphire substrate. For the active device, we used a commercial AlGaN/GaN heterostructure grown on a 430μm thick *c*-plane sapphire substrate. The epitaxial structure consists of a 3nm GaN cap, a 21nm AlGaN barrier, a 1nm AlN spacer, and a 1.8 μm GaN channel. A 2DEG naturally forms at the AlGaN/GaN interface. For the active sample, the first fabrication step was mesa isolation, which was performed using a $BCl_3/Cl_2$ plasma etch at 100 W to a depth of approximately 100 nm. This resulted in a mesa isolation width of approximately 10 μm. After removing the resist, we deposited a very thin $SiO_2$ layer (~3 nm) by atomic layer deposition (ALD) at 200°C. This layer helps raise the breakdown voltage. We patterned a bilayer resist to form the interdigitated Schottky contacts, giving an undercut profile needed for lift-off. A Ni/Au stack (30 nm/270 nm) was then deposited by e-beam evaporation, and the excess was lifted off in acetone. To end, the device was passivated with a thin $SiO_2$ film (~30 nm) using RF sputtering. The final device featured a grating structure with fingers of width 200 μm that overlapped on the mesa region approximately 3.5 μm on either side, leaving a gap of about 3 μm. This overlap also weakens the concentrated peak electric field, preventing premature breakdown. For comparison, we also fabricated a reference device with the same grating layout on bare sapphire, but in this case, only photolithography and Ni/Au metallization steps were performed.

The devices' frequency-domain transmission characteristics were measured with a commercial fiber-coupled spectrometer (TOPTICA TeraScan) at room temperature. As shown schematically in **Fig. 2(a)**, a linearly polarized THz beam (TE polarization) was focused onto the device at normal incidence. We performed a broadband (0.2–0.6 THz) transmission measurement of both the passive reference device and the active AlGaN/GaN device at zero bias. Furthermore, the active device was characterized in the 0.51–0.58 THz range by sweeping a DC bias of 0 to 35 V across adjacent fingers, which tunes the depletion of the two-dimensional electron gas and drives the modulation response. The profound effect of the 2DEG depletion is shown in **Fig. 2(b)**, which compares the spectra of the passive and active devices at zero bias. In contrast to the resonant response of the passive device, with strong EOT transmission peaks at frequencies above 0.42 THz, the active device spectrum shows these features suppressed. Normalized transmission levels at these peaks drop from near 0.6 down to about 0.1. The resonance is suppressed because of the conductive 2DEG. At zero bias, the free carriers provide a high conductive path across the grating gaps, effectively turning the structure reflective. This reflective channel screens the resonant fields and prevents the buildup needed to sustain high-Q features[21].

**Fig. 2(c)** shows that our electromagnetic model agrees with the measured spectrum of the passive reference device. The model also captures the positions of the broad FP oscillations and the sharp high-Q guided-mode resonances. The narrow resonances (for example, at 478 GHz) slightly lower measured peak transmission likely comes from added losses due to fabrication imperfections (lithography and metal losses) as well as misalignments during measurements. This effect is more pronounced for the resonances with the largest quality factor, i.e., those with the narrowest spectral width[12].

**Fig. 2(d)** illustrates active modulation. The interdigitated structure acts as an array of back-to-back Schottky diodes[22]. When a voltage is applied between adjacent fingers, one junction is driven into forward bias while the other is reverse-biased. The reverse-biased junction sets the device's electrical response, which limits the current flow. As the reverse bias increases, the 2DEG beneath the biased finger starts to be depleted, and is completely depleted at the pinch-off voltage. Beyond the pinch-off voltage, the onset of lateral depletion begins in the ungated gap. It is observed that after a modest bias (above 10 V) is applied, the transmission amplitude begins to rise. This lateral depletion remains confined to hundreds of nanometers, even at large reverse bias. Such lateral depletion is significant, as it directly governs the degree of channel pinch-off and therefore the modulation efficiency of the device. It is important to note that the total capacitance of these AlGaN/GaN-based diodes includes components from both the vertical depletion under the contacts and the lateral depletion near the contact edge, along with parasitics[23,24]. Equation 3 allows for analytical estimation of the lateral depletion width [25],

$$x_{dep} = \frac{2\varepsilon_r \varepsilon_0 \left(V - V_p\right)}{qNs}, \tag{3}$$

Where $\varepsilon_r$ is the relative permittivity of GaN, $\varepsilon_0$ is vacuum permittivity, $V$ is the applied voltage, $V_p$ is the pinch off voltage, $q$ is the electron charge and $N_s$ is the sheet charge density. Using the sheet charge density extracted for the present 2DEG, Eqn. (3) yields $x_{dep} \approx 0.22\,\mu m$ for a 35 V bias voltage. This 1D approximation does not account for fringing fields and polarization, which further increase the depletion width.

At 35 V (**Fig. 2(d))**, the measured transmission peak reaches its maximum value; further increase in bias voltage does not improve the transmission amplitude. At 0.53 THz, the modulation depth at resonance is calculated to be about 72%, based on the definition $(T_{max}-T_{min})/T_{max}$. For comparison, the spectrum of the passive reference device is overlaid, representing the maximum achievable transmission if there was no 2DEG. The modulated transmission peak at 35 V does not fully recover to the level of the bare sapphire reference. This could be attributed to residual free carriers in the undepleted portion of the 2DEG, additional losses from the AlGaN/GaN epilayers, fabrication imperfections, misalignments during measurements, as well as metal losses in the metallic grating.

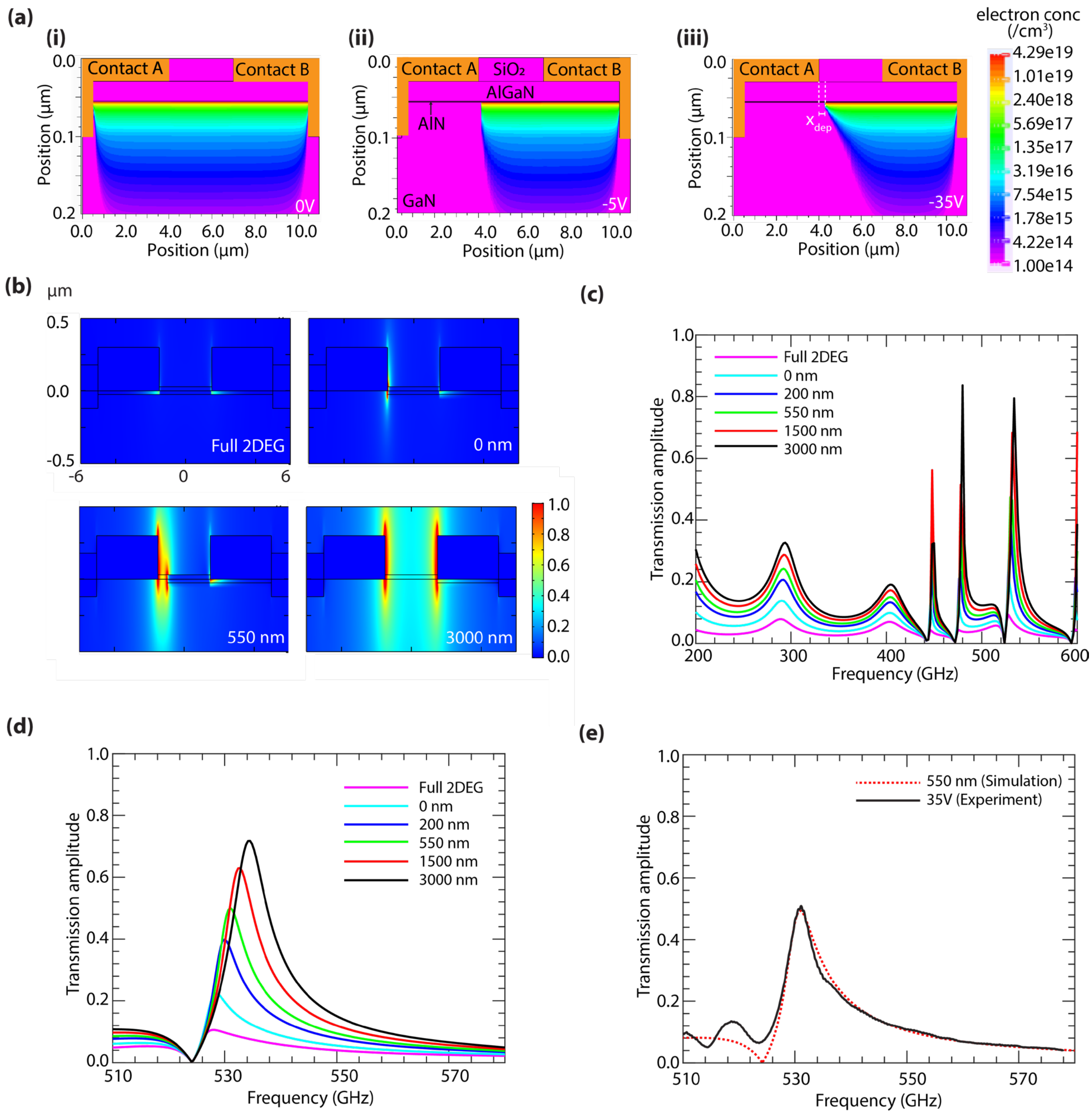

**Figure 3. (a)** Simulation using Silvaco TCAD, showing contour maps of electron concentration under voltage bias of (i) 0V, (ii) -5V, and (iii) -35V. **(b)** Simulation using COMSOL, showing a color map of the terahertz electric field distribution within the mesa and gap regions for various lateral depletion lengths. **(c)** Simulated transmission spectra for different lateral depletion lengths on the 200 to 600 GHz range. **(d)** Simulated transmission spectra focused on the 510–580 GHz frequency range. **(e)** Comparison between the simulated transmission spectra at a depletion length of 550 nm and the experimental transmission spectra measured at 35 V.

To gain a better understanding of the extent of lateral depletion, 2D device simulations were performed using Silvaco TCAD Software. To simplify the simulation, we model approximately 11 μm of width across the mesa region, with a gap of 3 μm, using the parameters stated before. **Fig. 3(a)** shows the contour maps of the electron concentration in the device under different voltage biases ranging from 0 to -35V. Under a bias voltage of 0V, see **Fig. 3(a)(i)**, there exists a high electron concentration across the whole channel. Under a bias voltage of -5 V (pinch off voltage in agreement with the results of CV characteristics, not shown here), see **Fig. 3(a)(ii)**, the 2DEG channel is depleted completely under contact A. Under -35V, the 2DEG depletion expands laterally by a few hundred nanometers, as shown in **Fig. 3 (a)(iii).** .

We also performed 2D full-wave electromagnetic simulations in COMSOL Multiphysics, modeling various lateral depletion widths. As shown in the field plot in **Fig. 3(b)**, the edge of the 2DEG is assumed to originate at the edge of the contact, where lateral depletion begins. In **Figs. 3(c-d),** it is observed that an increase in the depletion width results in an increase in the transmission amplitude, similar to what is observed in **Fig. 2(d).** From the simulations in **Figs. 3(a)(iii)** and **Fig. 3(d)** we observe that the observed experimental transmission results agree well with a lateral depletion of 550 nm (see **Fig. 3(e)**). These results verify that the depletion width at the maximum applied voltage is partial and extends a few hundreds of nanometers in agreement with our estimations from evaluating **Eqn. (3)** [25].

It is worth mentioning that while 2D longitudinal plasmons have been extensively demonstrated in GaN-based systems with sub-micron ungated gaps and metallic gratings[18,26–29], the spectral response observed in our setup is dominated by photonic FP interference and GMRs, which are determined by the grating geometry and electromagnetic boundary conditions. Similar to what has been observed by Shrekenhamer et al[4] (in this case, in metamaterial structures composed of split ring resonators), the plasmon instability by the 2DEG does not play a role in our response; the dynamic response of the carrier density in the channel is governed in our case by the frequency-dependent conductivity, exhibiting small influence from its imaginary part. Consistent with prior work from our group on GaN grating-based devices, plasmonic features do not significantly affect the 2DEG response at room temperature due to the short electron momentum relaxation time; however, they can become well-defined at cryogenic temperatures[19]. The modeling approach employed in this work is similar to that in our previous work; therefore, if plasmonic modes were present in the current structure, they can be captured by the model.

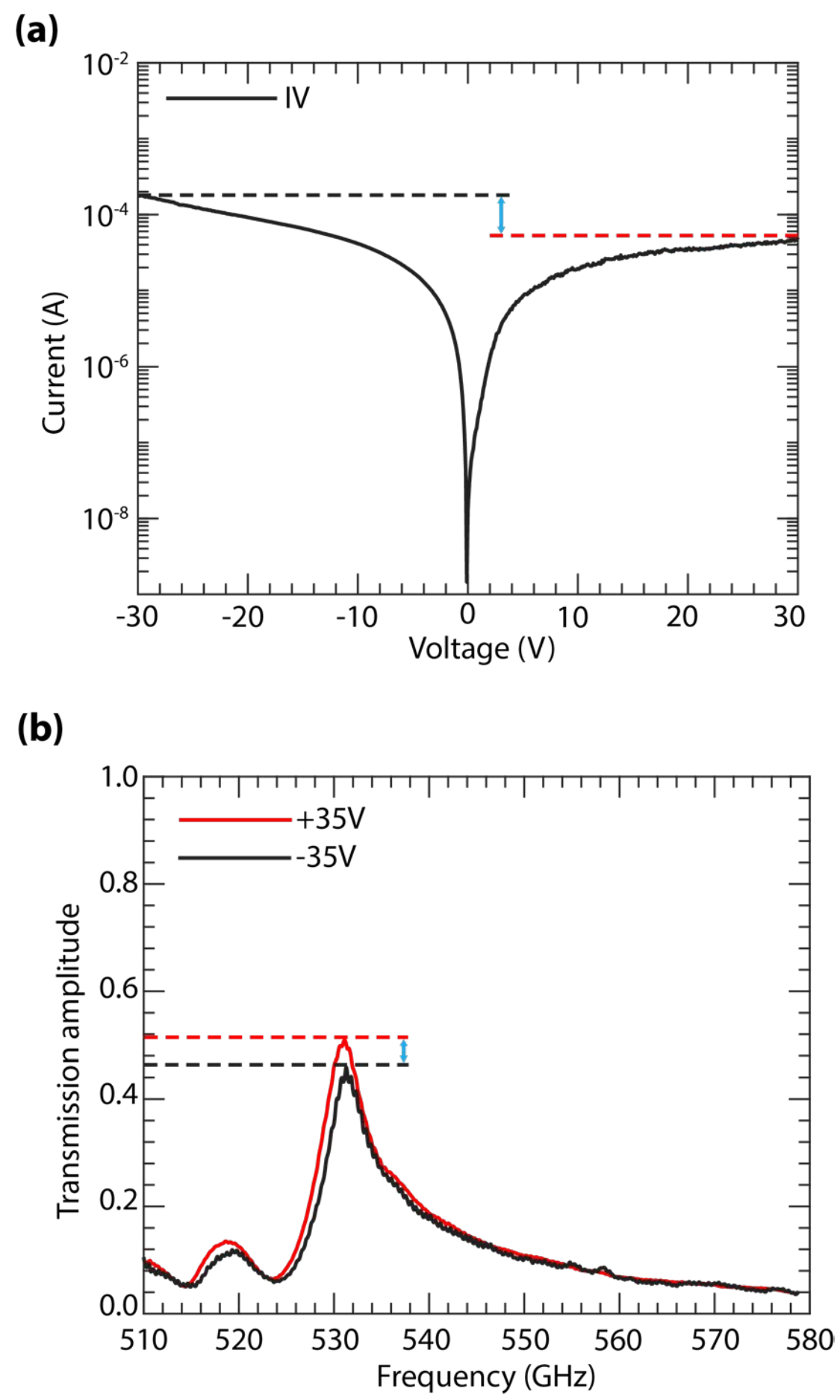

**Figure 4. (a)** Current-voltage (I-V) characteristic of the device. The distinct asymmetry indicates that the back-to-back Schottky contacts have unequal barrier heights, a result of non-uniformities at the metal-semiconductor interface. **(b)** Corresponding THz transmission spectrum. The difference in the barrier heights leading to the I-V asymmetry shown in (a) results in the device's bias-asymmetric THz response.

The devices' current-voltage (I-V) characteristics were measured using a Keithley 4200A-SCS Parameter Analyzer. The I–V curve in **Fig. 4(a)** is clearly asymmetric, suggesting that the two back-to-back Schottky contacts have different barrier heights. This effect can still arise even if the contacts are made under identical conditions, because of minor, uncontrollable variations at the metal–semiconductor interface[22]. The Schottky barrier height is a key factor for DC leakage and affects the THz modulation. A smaller barrier leads to much higher reverse leakage current and reduces how effectively the 2DEG channel can be depleted. Given this asymmetry, the device's modulation was characterized by sweeping the bias in both polarities to demonstrate the asymmetry of depletion effect on THz transmission clearly. This difference in the electrical barriers is directly reflected in the THz transmission response shown in **Fig. 4(b)** with larger transmission attainable for positive voltages.

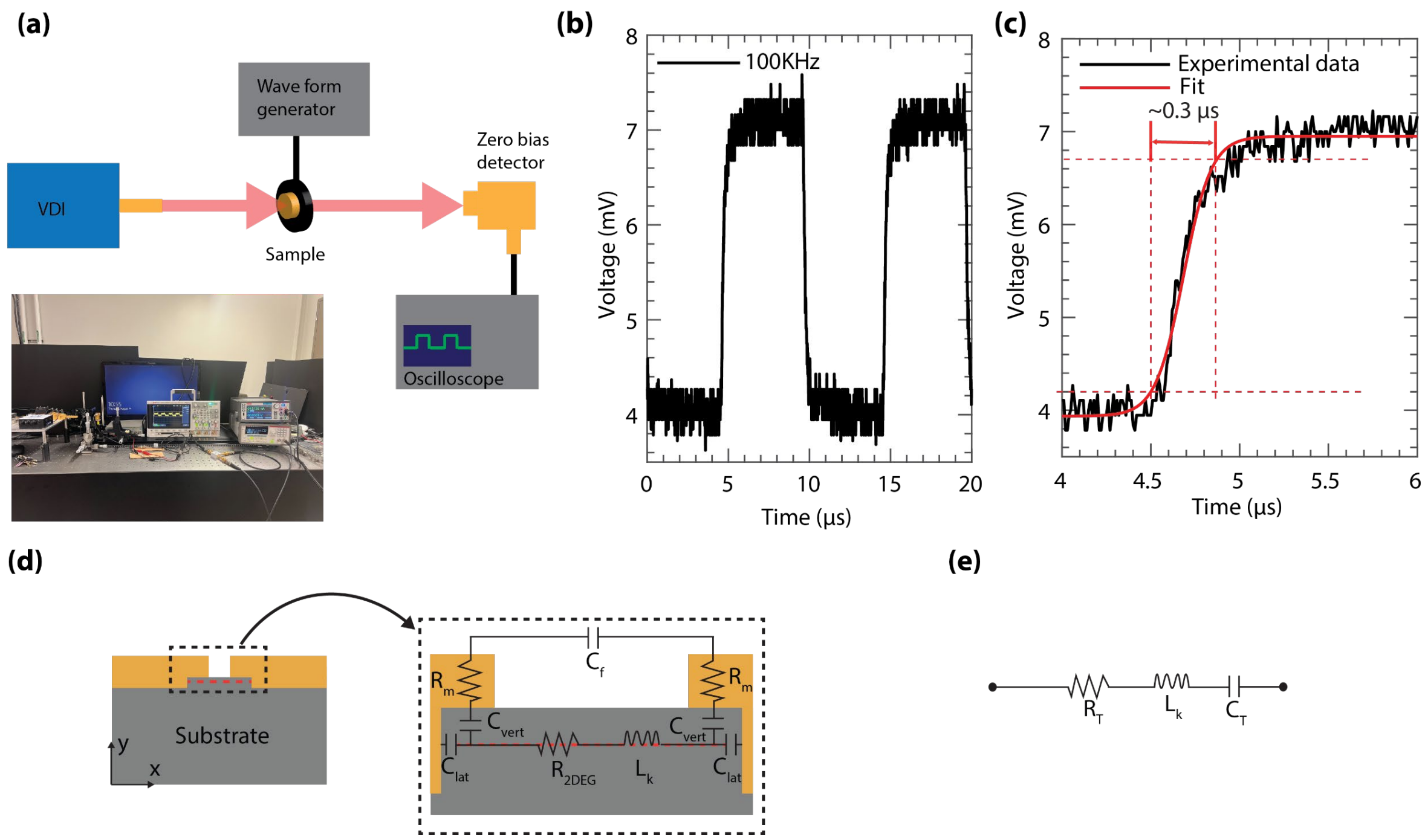

**Figure 5**. Dynamic characterization. **(a)** Schematic and optical image of the measurement setup. **(b)** Modulated THz signal at the detector in response to a 100 kHz square-wave input. **(c)** Magnified view of the rising edge from (b), indicating a 10–90% rise time of approximately 0.3 µs. **(d)** Schematic of a unit cell with equivalent circuit. **(e)** small signal equivalent circuit.

The device dynamic response was measured using a 0.3 THz continuous-wave (CW) multiplier-based THz source from VDI, a zero-bias detector (ZBD), and an oscilloscope as illustrated in **Fig. 5(a)**. We drove the grating with a square-wave voltage consisting of an AC signal of 10 V (peak-to-peak) riding on a DC bias of 15V that depletes the channel. The modulator's time-response (signal detected by the ZBD) is plotted in **Fig. 5(b-c)**. The oscilloscope trace shows a 10–90% rise time of about 0.3 µs, corresponding to a cutoff frequency of roughly 1 MHz. The extrinsic RC time constant of the large-area device architecture governs this measured speed. Note that the modulation depth is about 43%; the drive voltage in this case swings from 10 to 20V, and the source frequency is 0.3 THz, which corresponds to a spectral region where the response is dominated by FP resonances; this is different from the modulation attained at the maximum resonance region of 0.53 THz. We estimated the intrinsic cutoff frequency from the device's total series resistance and junction capacitance. When about ~15 Ω resistance from the metal components (such as the bus bar) is considered, the intrinsic speed reaches roughly 220 MHz. The discrepancy between the measured bandwidth and the intrinsic device speed is due to significant resistance and capacitance (RC constant)[30–32]. These parasitics arise from series resistance in the forward-biased Schottky contact, undepleted 2DEG channel, metallic traces, parasitic inductance from device packaging and interconnects (e.g., wire bonds, metallic contacts), and parasitic capacitances from the interdigitated fingers and contact pads. This RC limitation can be analyzed using the unit cell and the small-signal equivalent circuit of the back-to-back Schottky diode shown in the **Figs. 5(d, e)** respectively. The total capacitance, $C_T$ is composed of $C_{vert}$, $C_f$, and $C_{lat}$. Where $C_{vert}$, $C_{lat}$, and $C_f$ refer to the vertical capacitances of the dielectric layers above the 2DEG layer, the lateral capacitance, and the fringing capacitance between the fingers in the channel, respectively. The total resistance, $R_T$ consists of $R_m$ and $R_{2DEG}$, which refer to the resistance of the Schottky metal and the channel resistance, respectively. Finally, $L_k$ represents the kinetic inductance arising from electron inertia [33]. At zero bias, the capacitance of the diode is dominated by the vertical capacitance of the dielectric, since $C_f$ is negligible. As the bias increases beyond the pinch-off voltage, the capacitance sharply decreases and becomes dominated by the small capacitance of the laterally depleted region, which governs speed. Although the performance is below the intrinsic limits of GaN, this first-generation prototype successfully demonstrates that the active free-space modulator is a viable and effective mechanism for THz modulation. For context, AlGaN/GaN-based terahertz modulators reported in the literature exhibit a wide range of performance characteristics that depend on the underlying modulation mechanism. For example, optically driven grating-gate plasmonic structures have exhibited ultrafast modulation dynamics with relaxation times on the order of 15–20 ps[18], electrically driven 2DEG-based metasurface modulators have achieved large modulation depths of up to 85% and electrical modulation speeds of up to the gigahertz regime, primarily through broadband or non-resonant operation[17]. In resonant grating-coupled plasmonic modulators, modulation depths above 90% have been reported, most often under cryogenic conditions and with limited modulation bandwidth[16]. In the present device, modulation is realized through electrical control of EOT via GMR. This provides narrow frequency-selective modulation with high-Q resonances at room temperature that are particularly attractive for terahertz communication and sensing applications that require narrow-band operation. Future work will focus on enhancing device speed by minimizing parasitics and reducing the active area, thereby reducing the overlap region and enabling GHz-class operation.

## IV. CONCLUSIONS

In summary, we have realized an electrically tunable terahertz modulator based on an interdigitated Schottky grating integrated with an AlGaN/GaN heterostructure. The applied bias laterally depletes the two-dimensional electron gas, enabling dynamic control of coupling between the incident field and guided-mode resonances in the substrate. This change in charge distribution governs the strength of the extraordinary optical transmission observed near 0.53 THz. The measured spectra show excellent agreement with full-wave electromagnetic simulations, validating the resonance-based modulation mechanism. The structure achieves a maximum modulation depth of 72% and a switching speed with associated cut-off frequency of approximately 1 MHz, demonstrating that lateral carrier control provides an efficient means to tune resonant field coupling in the terahertz regime.

## Acknowledgments

This work was supported by the University of Utah research foundation under seed grant "Advancing High-Speed THz Modulators with Lateral GaN Junctions". This material is also based upon work supported by the Air Force Office of Scientific Research (AFOSR) under Award No. FA9550-18-1-0332. Any opinions, findings, and conclusions or recommendations expressed in this material are those of the author(s) and do not necessarily reflect the views of the United States Air Force. This work was performed in part at the Utah Nanofab sponsored by the College of Engineering, Office of the Vice President for Research, and the Utah Science Technology and Research (USTAR) initiative of the State of Utah.

## Author Declarations:

### Conflict of Interest

The authors have no conflicts to disclose.

## Author Contributions

**Geofrey Nyabere:** Conceptualization (equal); Data curation (lead); Formal analysis (lead); Investigation (lead); Methodology (lead); Visualization (equal); Writing–original draft (lead); **Hunter Ellis:** Investigation (supporting); Methodology (supporting). **Miguel Gomez:** Investigation (supporting); Methodology (supporting). **Wei Jia:** Investigation (supporting); Methodology (supporting); Formal analysis (supporting); Writing – review & editing (supporting). **Yizheng Liu:** Investigation (supporting); Methodology (supporting). **Karli Ann Higley:** Data curation (supporting); Investigation (supporting). **Sriram Krishnamoorthy:** Investigation (supporting); Methodology (supporting). **Steve Blair:** Investigation (supporting), Methodology (supporting); Formal analysis (supporting); Writing – review & editing (supporting). **Kai Fu:** Funding acquisition (supporting); Resources (supporting); Formal analysis (supporting); Methodology(supporting); Writing – review & editing (supporting). **Berardi Sensale-Rodriguez:** Conceptualization (equal); Funding acquisition (lead); Supervision (lead); Formal analysis (equal); Project administration (lead); Resources (lead); Writing -original draft (equal); Writing – review & editing (lead).

## Data Availability

The data that support the findings of this study are available from the corresponding author upon reasonable request.